\documentstyle[12pt,aasms4]{article}
\chardef\isp="10
\def\i{\'\isp}
\lefthead{Jaffe and Mart{\i}n--Pintado}
\righthead {Low Velocity Ionized Winds}
\begin{document}
\title{Low Velocity Ionized Winds from Regions Around Young O Stars}
\author{D.T. Jaffe}
\affil{Department of Astronomy, University of Texas at Austin}
\authoraddr{Austin, Texas 78712; dtj@astro.as.utexas.edu}
\author{J. Mart{\i}n--Pintado}
\affil{Instituto Geogr\'afico Nacional, Observatorio Astron\'omico Nacional}
\authoraddr{Apartado 1143, E-28800 Alcal\'a de Henares, Spain; martin@oan.es}
\def\ga {$\alpha$}
\def\kms {km s$^{-1}$}
\def\uc {^}
\begin{abstract}
We have observed seven ultracompact H~II regions in hydrogen
recombination lines in the millimeter band.  Toward four of these 
regions,
there is a high velocity (full width to half maximum 60--80 \kms)
component in the line profiles.  The high velocity gas accounts
for 35\%--70\% of the emission measure within the beam. We compare
these objects to an additional seven similar sources we have found
in the literature.  The broad recombination line objects (BRLOs)
make up about 30\% of all sources in complexes containing
ultracompact H~II regions.  Comparison of spectral line and continuum
data implies that the BRLOs coincide with sources with rising spectral
indices, $\geq$0.4 up to 100 GHz.  Both the number of BRLOs and
their frequency of occurrence within H~II region complexes, when coupled
with their small size and large internal motions, mean that the 
apparent contradiction between the dynamical and the population lifetimes
for BRLOs is even more severe than for ultracompact H~II regions
as a whole.  We evaluate a number of possible models for the origin
of the broad recombination line emission.  The lifetime, morphology, and rising
spectral index of the sources argue for photoevaporated disks as
the cause for BRLOs.  Existing models for such regions, however,
do not account for the large amounts of ionized gas
 observed at supersonic velocities.

\end{abstract}
\keywords{ISM:H~II Regions ISM:Jets and Outflows Radio Lines:Stars
  Stars:Mass Loss}

\section{Introduction}
The formation, behavior, and evolution of O stars influence many
aspects of the interstellar medium in the Galaxy.
O stars play a central role in the evolution of galactic 
abundances, influence the structure of the interstellar medium
through supernova explosions, ionization, and mass-loss,
 and determine the appearance of the 
spiral arms in the visible and the radio through both thermal and
non--thermal emission.  
The pre--main sequence evolution of luminous O 
stars takes place while they are still shrouded within
their natal clouds. 
Therefore, we usually 
study the surrounding H~II regions and dust clouds, rather than the 
stars themselves, and carry out these studies at far-IR and radio 
wavelengths.

Much of the gas in the H~II regions that surround newly formed O stars 
is involved in supersonic motions of the order of 20 \kms,
usually ascribed to ``turbulence''.  
The smallest and presumably youngest H~II regions (ultracompact, or 
UC H~II regions) typically have linewidths of 25--35 \kms\ in high frequency
centimeter wave recombination lines (RRL's, Garay 1989, Churchwell et al. 1989,
Afflerbach et al. 1996).
There is considerable observational evidence for even higher velocity
motions of various sorts.
Massive O stars have winds with terminal velocities of
1--2$\times$10$\uc3$ km s$^{-1}$ and mass loss rates of up to a few 10$^{-6}$ 
M$_{\sun}$ per year (Chlebowski \& Garmany 1991).  
Many high luminosity and presumably young stars embedded deep within
molecular clouds show evidence for ionized circumstellar envelopes
or for winds with terminal velocities $\sim$200 \kms\ (Simon et al. 1981, 
1983, Persson et al. 1984, Garden \& Geballe 1986, Bunn, Hoare, \& Drew 1995).
In some more evolved and optically accessible H~II regions, there are small
amounts of gas moving at 100-200 \kms\ with respect to the bulk velocity
of the H~II region (Meaburn \& Walsh 1981, 1986).

There is increasing evidence that a (substantial) minority of ultracompact
H~II regions emit  broad radio recombination lines. 
(In this paper, we refer to these sources as broad recombination line
objects, BRLOs).
The presence of this RRL component implies that the regions
contain gas moving at velocities up to $\pm$60 \kms\
from the rest velocity of the cloud (Altenhoff, Strittmatter, \& Wendker 1981,
Zijlstra et al. 1990, Gaume, Fey, \& Claussen
1994, De Pree et al. 1994, 1996).  In the UC H~II regions, unlike the more
evolved H~II regions, the high velocity gas may account for a significant
fraction of the total emission measure.  
If high velocity emission arises in the very compact parts of the UC H~II
region complexes where the emission measures are high, the gas may only 
become detectable in observations at the shortest
centimeter wavelengths where these compact sources 
begin to become optically thin and where non-LTE effects become negligible.  
Observations at shorter wavelengths could therefore provide a better way
to search for the broad features.
We have investigated the incidence
and nature of the broad recombination line features by surveying
a sample of bright UC H~II regions in 3, 2, and 1.3 mm recombination
lines.  This paper presents results and analysis of this survey and
a discussion of the origin of the broad line emission.

\section{Observations}

We made all the observations we present here 
with the IRAM 30 meter telescope at Pico de 
Veleta (Spain).  The first set of observations took place in 1992
February and the second set in 1993 March. 
On the fist observing run, we observed three recombination lines simultaneously 
at wavelengths of  3, 2 and 1.3~mm, and on the second run, 
we observed several combinations of two such lines (see Table \ref{tab1}). 
The telescope beam at these  wavelengths was, 
respectively, 24, 17 and 11 arc seconds.
Table 1 lists the sources, source positions, dates of observations,
and observed transitions. The receivers were SIS mixers, tuned to single 
sideband 
(image rejections between 7 and 10 dB), with typical system noise 
temperatures of 250, 350 and 650 K at 3, 2 and 1.3 mm. 
We used two 512 
channel filterbanks with 1 MHz resolution and a 430 channel Acousto-Optical 
Spectrometer  with a  resolution of 1.1 MHz as spectrometric backends. 
In order to insure good baseline subtraction, we nutated
the telescope subreflector at 0.25 Hz between
the source and a reference position 240 arc seconds away.  While some of the 
sources have extended radio continuum emission, 
published maps at centimeter wavelengths show that an angular
displacement of this size should result in contamination of the
reference beam by narrow recombination line emission at only a very low level. 
Under good weather conditions, use of the nutating subreflector 
allows one to measure the line and continuum emission simultaneously. 
The weather conditions during the 1993 observations were good enough
to permit us to measure the
continuum emission at 3~mm.
We calibrated the data by observing the atmosphere and   
cold and hot loads at known temperatures. 
All the intensities are in units of 
antenna temperature corrected for forward scattering and spillover as well
as for atmospheric transmission, T$_A\uc*$, and the conversion 
factor from antenna temperature 
to flux density at 3~mm is  6 Jy/K.


Almost all of the ultracompact H~II regions studied as part of the
present work lie close to or within dense, high column density 
molecular cores.  These cores produce very rich millimeter wave line
spectra.  Confusion from these molecular line emitters within the
beam is an important limiting factor when deriving quantitative
results from the millimeter recombination line data. The presence of
the molecular lines does not, however, impede our ability to {\it detect}
high velocity recombination line emission.  The molecular 
lines are, in general, quite narrow (5--8 km s$^{-1}$) and the broad
components of 
adjacent $\Delta$n=1 recombination transitions with similar n should be similar
in strength and shape but overlap with completely different sets
of molecular features.

\section{Results and Analysis}

Table \ref{tab2} presents the best fits of one or two Gaussians 
to profiles of selected lines in each source in our survey.
In all cases, the lines selected from among the 1.3--3~mm Hn$\alpha$
transitions observed (Table \ref{tab1}) are the 3~mm transitions,
H39$\alpha$ and H41$\alpha$. Since our observed positions coincide with
or lie close to hot, dense, high column density molecular cores 
(cf. Plume, Jaffe, \& Evans 1992, Plume et al. 1997),  
the high density of molecular features at 2~mm and 1.3~mm makes the
hydrogen recombination lines in these bands more prone to confusion.
The 3~mm continuum flux densities measured toward
the positions given in Table \ref{tab1} also appear in Table \ref{tab2}. 
Table \ref{tab2} gives
the electron temperatures,
derived from the ratio of the total integrated H41$\alpha$ line intensity 
(including both line components if more than one is present)
to the measured continuum adjacent to this 3~mm line, assuming LTE conditions.
It is interesting to note that failure to notice the 
presence of a relatively weak broad component in recombination line spectra,
a component which can be missed in 
observations with moderate sensitivity, can lead one
to overestimate the electron 
temperatures derived from the line-to-continuum ratios. 
The LTE electron temperatures we derive from our
data all lie within $\pm$30 \% of the canonical 10,000 K temperature of
H~II regions.    

\subsection{Broad Recombination Line Objects}

The millimeter recombination line shapes for four of the seven sources
in our sample (NGC 7538, W49N, G34.3+0.2, and S106)
 deviate substantially from the simple 25--35 \kms\
wide Gaussian profiles that are typical of ultracompact H~II
regions.  Fits of single Gaussians to these profiles left systematic
residuals at $|V|\geq$20 \kms\ from the line center.  
In all cases, we significantly reduce the residuals when we fit
the observed spectra with
two Gaussian profiles, one with the normal UC H~II region recombination
line width and another broad component which is 2--3 times wider.
We present here our analysis of these four sources.

Toward most of the sources in our sample, 
the 30~m beam contains very complex radio continuum structure that contributes
to both the normal and the broad spectral components.  
So far, there are no maps
of recombination lines at millimeter wavelengths with sufficient angular
resolution and sensitivity to identify the sources responsible for the
broad features.  To identify the sources giving rise to these features, we 
combine our millimeter recombination line
data with existing high angular resolution maps of continuum and recombination
line
emission, mainly observed at centimeter wavelengths. 
To aid in comparisons with these maps, we
estimate the amount of continuum 
emission ``required'' to produce each line component
by assuming  that the observed 3~mm line and continuum emission is 
optically thin and that the millimeter
recombination lines arise
from gas where the  electron temperature equals
the temperature derived from the total line and continuum emission
(under the assumption of LTE conditions)  
from each source (Table \ref{tab2}).
Table \ref{tab2} gives our estimate of the
continuum associated with the broad and the normal components based on these
assumptions.  
The gas that produces the broad line component accounts for 35\%--70\% of the 
3 mm continuum flux and therefore at least that fraction of the emission
measure in these four objects.

\subsubsection{ NGC 7538}

We have observed six millimeter wave hydrogen recombination lines
toward NGC 7538.  
Our beam center lies at the position
of the infrared source and double radio continuum source NGC 7538 IRS1
(Campbell 1984). A free fit of a single Gaussian profile to the H41$\alpha$
line underestimated the brightness both at the peak of the line and
at high velocities.
The left panel of
Figure \ref{n7538comp} shows the results of a fit which we forced to
match the area around the peak by minimizing the  least squares deviation
between a single Gaussian profile 
and the observed line over a range of $\pm$15 \kms\ about the peak of
the line.  The lower part of the
 left panel shows that, while this profile accounts
well for the inner part of the observed line, there are substantial residuals
to this fit, in
particular at higher velocity.  The right
panels of Figure \ref{n7538comp} show a fit of two Gaussians to the
same profile and the residuals resulting from this fit. 
  The line parameters we derive in an
independent fit of two Gaussians to the 
H39$\alpha$ line agree well with the H41$\alpha$ results (Table \ref{tab2}).
For both hydrogen lines, the normal line has a width of around 25 \kms\
while the broad component has a width of 70 \kms.

High resolution 1.3 cm recombination line observations toward NGC 7538
IRS1 (Gaume et al. 1995) reveal extremely broad emission (full width
to zero power of around 200 \kms\ in the northern and southern
hypercompact lobes and an absence of line emission in between the lobes.  
 The discrepancies between the linewidths of the centimeter and millimeter 
wavelength recombination lines are similar to those found in MWC~349. As in 
this latter  source, the broad H66$\alpha$ lines in NGC~7538~IRS1, 
 which show  noticeable 
non-Gaussian profiles with flat--topped shapes, are very likely affected by 
non-LTE excitation (see Mart{\i}n-Pintado et al 1993).

 For NGC 7538, our beam contains the sources IRS1, IRS2 and IRS3 and the total 
continuum flux density at 3~mm  is in good agreement with the fluxes derived by 
Akabane et al. (1992)  from  interferometric measurements. 
The 3$\arcsec$ resolution 3~mm continuum map (Akabane et al. 1992) shows that 
IRS3 makes a negligible contribution to the millimeter wave continuum 
flux, IRS2 contributes 1 Jy, and IRS1 
contributes 1.6 Jy.  The values for the continuum fluxes for the two
latter sources are consistent with an attribution of the normal spectral
component to IRS2 and the broad component to IRS1, the same sense in
which the origin of the normal and very broad 6 cm recombination line emission
is directly attributable to the respective sources (Gaume et al. 1995).

\subsubsection{ W49N}

We observed six millimeter wave hydrogen recombination lines
toward this source (Table \ref{tab1}).  
Our beam center lies toward the H$_2$O maser center,
about 1\arcsec\ north of the ultracompact 15 GHz
continuum source
G1 (Dreher et al. 1984).  
The upper left panel of Figure \ref{w49n} shows a single Gaussian fit to the
H41$\alpha$ line made following the procedure described for the single
Gaussian fit to the NGC 7538 profile.
The residuals
to this fit (lower left) show a systematic excess in the wings of the
line.  The upper right panel of Figure \ref{w49n} shows a two Gaussian
fit to the same line.  The minimization results in a component with
a velocity width of 31 \kms\ and a second component 71 \kms\ wide.
Over the velocity interval occupied by the hydrogen line, 
the residuals to the two Gaussian fit at the central position
are flat. 
We also fit two Gaussians to the He41$\alpha$ line at this position.
We allowed the temperatures of the two components to vary but fixed
the widths of the two He components at the values of the
H41$\alpha$ widths and the velocity offset from the corresponding H41$\alpha$
components at --122.2 \kms. This fit (Figure \ref{w49n} (lower right))
shows that a normal plus a broad component can account
for the shape of the helium line as well. 
The strengths of the normal and broad
He components are respectively 9\% and 8\% of the strengths of the  
corresponding hydrogen emission components.
A five point map of the H41$\alpha$ transition at 12$\arcsec$ intervals
is consistent with an unresolved source as the origin of both the
normal and the broad component.

  The upper panel of Figure \ref{w49compadd} superposes
spectra of H39$\alpha$
and H41$\alpha$ toward W49N. The wings of the two recombination lines
match very closely, but the shapes of the inner portions of the lines
differ considerably. 
A triple Gaussian fit to the H39$\alpha$
spectrum fixing the velocities and widths
of the two broader components to match the values derived from
the H41$\alpha$ spectrum shows that there is a third feature with a 
full width to half maximum of 13.3 \kms.
This narrower feature has a width
similar to the width of molecular lines in this source (Serabyn, Guesten
\& Schulz
1993). The third line has an observed velocity of --7.5 \kms\ 
with respect to the H39$\alpha$ frequency.
The only molecular transition with a rest frequency close to that of the
H39$\alpha$ line (106737.365 MHz) is the $2_3-1_2$ transition of $\uc{34}$SO
($\nu_o$=106743.365 MHz).  For this line, the observed velocity 
corresponds to V$_{LSR}$= 6.6 \kms\ in its own frame.
This velocity is close to the typical molecular velocity of the W49 core.
The lower panel of Figure \ref{w49compadd} shows an overlay of the
observed H39$\alpha$ and H41$\alpha$ profiles 
where we have added the fitted $\uc{34}$SO feature to
the H41$\alpha$ data to illustrate the similarity of the shapes of the 
recombination line profiles themselves.

There is some evidence for high velocity recombination line emission in
previous millimeter observations of W49N. The wings seen in Figure \ref{w49n}
are also apparent to the trained eye in the H40$\alpha$ spectrum
in Figure 3 of Gordon (1989).  High angular resolution observations of
the H66$\alpha$ and H52$\alpha$ transitions at 1.3~cm and 0.7~cm, respectively
(De Pree, Mehringer, \& Goss 1997), 
show that sources A and B, both of which lie 
within our beam, produce lines with a full width to half maximum of 
50-60 \kms. Since the line fills the spectrometer band for these
sources, however, 
the observations do not exclude the presence of higher velocity
emission consistent with our millimeter measurements.

Determining the source or sources in W49N responsible for the broad
line emission is complex because of the large number of 
continuum sources in our beam. 
The  H41$\alpha$ beam contains the sources A--G2.
We can use  high angular resolution maps of the 
RRL and continuum  
emission (De Pree et al. 1997) to establish 
which source may be 
responsible for the broad component. As previously
mentioned, sources A and B have the
broadest H66$\alpha$ and H52$\alpha$ profiles. 
These two sources 
have power law radio continuum spectra with spectral indexes of 0.6 and 
0.9, respectively. Extrapolating to 3 mm, A and B would each  contribute 
around 2 Jy. Given the 3 Jy required to provide the source of the
broad line emission (see Table \ref{tab2}), these two
stellar wind sources are viable candidates for the origin of the broad lines.
However, source A is 
located at the edge of the beam and the map of the 
H39$\alpha$ line indicates a point 
source  located closer to B than to A. Like in
NGC 7538 IRS1, the broad component seems to 
be associated with compact sources with rising continuum spectra.

\subsubsection{ G34.3+0.2}

We observed six millimeter wave hydrogen recombination lines
toward this source (Table \ref{tab1}).  
The center position for these observations 
lies about 3\arcsec\ south of the
15 GHz continuum peak (Garay, Rodriguez, \& van Gorkom 1986).  
The left panel in Figure \ref{w44} shows a single Gaussian fit to the
H41$\alpha$ transition toward G34.3+0.2 (see the discussion of NGC 7538
for a description of the fitting procedure).
The residuals (lower left) show significant excess emission at high 
velocities, particularly on the blue side of the line.  The right panel
of Figure \ref{w44} shows a fit using two Gaussians.  Superposed on
the residuals of this fit is a fit to the helium line made by holding
the relative velocity to the corresponding hydrogen component fixed
and setting the widths of the two helium components equal to those of
the hydrogen components.  Figure \ref{w44comp} shows a superposition
of the H41$\alpha$ and H39$\alpha$ lines toward G34.3+0.2.  The figure
also shows the sum of two Gaussians with parameters averaged between
those from the H41$\alpha$ and H39$\alpha$ line fits. The compromise
Gaussians have widths of 32 and 75 \kms.  The shape of the
two hydrogen lines agree well, except at velocities where narrow molecular
lines from the hot core contaminate the profiles.

Toward G34.3+0.2, our 30~m beam at 3~mm contains three radio continuum sources.
Components A and B
are compact objects with spectra
that rise with frequency at least to 90 GHz (Gaume et al. 1994). 
Component C is a classic cometary
HII region (Reid \& Ho 1985; Garay et al. 1986).
At centimeter wavelengths, where sources A and B are weak and opaque,
all of the recombination line emission comes from
source C. 
The 2~cm results (Garay et al. 1986)
 indicate a relatively uniform line width of
around 50 \kms\ across source C.  The more sensitive 3~cm data (Gaume et al.
1994)
show line widths which increase from 25--45 \kms\ 
in the north and west to about 80 \kms\ in the southeast of component
C (Gaume et al. 1994). 

Our total continuum  emission measured at 
3~mm is $\sim$7 Jy, in good agreement with the 6.6 Jy we extrapolate
from the 6 and 2~cm measurements of Benson \& 
Johnston (1984) by assuming that  source C is optically thin and 
ultracompact sources A and B are optically thick with 
spectral indexes of  $\sim$0.9
(Gaume at al. 1994). Based on this extrapolation, the compact sources
A and B, despite their rising ``stellar wind'' 
spectra, can only account for 0.8 Jy of the 7 Jy flux density at 3~mm.
Both the broad and the narrow line component therefore should arise mainly
from the cometary H~II region.  
The radial velocities and widths of both the broad and narrow line
components in the H41$\alpha$ spectrum lie within the range of one
or another centimeter RRL measurement toward the parts of source C
that emit most of the flux (Garay et al. 1986, Gaume et al. 1994).
The results obtained by the centimeter RRL observers, however,
are not consistent with each other.

\subsubsection{S106}

S106 is less luminous than the other BRLOs
in our sample.  We selected this source
for investigation because of the numerous observations of broad
near-IR hydrogen recombination lines toward S106~IR (Persson et al. 1984,
Felli et al. 1985, Garden \& Geballe 1986, Drew, Bunn, \& Hoare 1993).
This IR source lies at the center of an obscuring dust disk
(Eiroa, Elsaesser, \& Lahulla 1979) and at the edge of our beam, 
13$\arcsec$ northeast
of the position listed in Table \ref{tab1}.  Coincident with the IR source
is a compact radio source with a spectrum rising as $\nu\uc{0.6}$
from 5 to 23 GHz (Snell \& Bally 1986, Felli et al. 1985).
Very high angular resolution 5 GHz observations show that this 
central source is elongated along the equator of the extended
bipolar optical and radio nebula (Hoare et al. 1994, Felli et al. 1984).
There is a strong submillimeter dust continuum source close to the
position we observed (Richer et al. 1993).  The difference between
the observed position and the IR source position and the presence
of relatively strong dust emission make modeling this source problematic.

For S106, the broad component cannot arise only from the compact continuum
source with the rising spectrum since
the extrapolation of the continuum spectrum of this source to 3~mm implies
that it will only account for 42 mJy at this wavelength.
 Therefore, the high-velocity recombination line
emission is associated with the
bipolar nebula.  Like G34.3+0.2, this source
appears to be more evolved than the UC HII regions in 
NGC~7538~IRS1 and W49N. Long slit spectroscopy in the optical shows
that S106 has the velocity structure of a bipolar outflow (Solf \& 
Carsenty 1982). There is  H$\alpha$
emission throughout the southern lobe of the extended bipolar source.  
H$\alpha$ has a total extent of over 100 \kms\ at the position we observed
at 3~mm.  Farther south in this lobe, the H$\alpha$ lines split into two
components with typical separations of 50 \kms\ (Solf \& Carsenty 1982).

\subsection{Narrow Line Sources}

\subsubsection{Orion/IRc2 and Orion H$_2$ Peak 1}

Figure \ref{oria} shows an H39$\alpha$ spectrum toward H$_2$ Peak 1 
({\it left}) and an H41$\alpha$ spectrum toward IRc2 ({\it right})
in Orion.  In both cases, the residuals below
each spectrum demonstrate that a single Gaussian fits the line profile well.
Hasegawa \& Akabane (1984) have reported high velocity emission
toward Peak 1 in the H51$\alpha$ line.  
The reported high velocity emission gave rise to the suggestion
that such emission could come from gas ionized by the radiative
precursor of a dissociative shock in the outflowing molecular gas 
(McKee \& Hollenbach 1987).
Our H39$\alpha$ observation at the position of Peak 1 excludes the
presence of a broad component at a level about a factor of five lower 
than the 0.1 K implied by the putative H51$\alpha$ detection.
Our H41$\alpha$ observations toward IRc2 also show no evidence for
emission beyond that arising from a single 23 \kms\ wide Gaussian
component. There is therefore no observational evidence
for shock-ionized gas in the region around Orion/IRc2.
It would be better to search for recombination
lines from shock-ionized gas where little or no confusing photoionized gas
is present.  Possible targets might include the high velocity 
shocked regions in powerful outflows associated with less luminous
sources like IRAS 03282 and L1448 (Bachiller 1996).

%
%

\subsubsection{G10.6-0.4}

G10.6-0.4 is a bright UC HII region situated in a molecular core.
Observations of the radio continuum at 6, 2, and 1.3 cm show several
emission knots within 5$\arcsec$ (Ho \& Haschick 1981, 1986).
We have observed two lines toward this source, H35\ga\ and H41\ga\
(Figure \ref{g106}). The single Gaussian fit to the
less confused H41\ga\ line  (Table \ref{tab2}) undershoots the peak
of the line and overshoots in the wings.  This behavior mimics that
of the residuals of single, unconstrained
 Gaussian fits to the lines in G34.3+0.2 and W49,
indicating the possible presence of a broad component.
The signal to noise ratio of the G10.6-0.4 spectra, however, is insufficient
to warrant a fit of more than one Gaussian.

\subsubsection{DR21}

We observed the H30$\alpha$, H35$\alpha$,
H36$\alpha$, H39$\alpha$, and H41$\alpha$ transitions
toward the compact H~II
region in DR21.  Figure \ref{dr21} shows the  H39$\alpha$ line.
The residuals in this figure indicate that a single Gaussiam
with a width of 32 \kms\ can account for all of the flux
in the hydrogen line.  This source has strong molecular and
neutral atomic emission at high velocities (Bally \& Lada 1983,
Russell et al. 1992) but does not have a corresponding ionized
gas component in the millimeter recombination lines.

\section{Other Broad Recombination Line Objects}


A search of the literature shows that there are at least seven additional
ultracompact HII regions with high velocity ($\Delta$V$\geq$55 \kms)
emission in recombination lines at n$\leq$80, where pressure broadening
is less likely to be important.  These sources are: the four sources comprising
SgrB2M F, G5.89-0.39, MWC~349, and K3-50A (see Table \ref{tab2}).

As a whole, 
{\bf SgrB2M} source F has broad H66$\alpha$ and H52$\alpha$ lines (Mehringer
et al. 1995, De Pree et al. 1996).  Integrating over the source, the
H52$\alpha$ line has a FWHM of 59 \kms.  The H66$\alpha$ observations
with the VLA resolve source F into four subsources.  
The line widths for these more compact objects range from 68 \kms\
to 79 \kms.
Each of the subsources
has a rising continuum spectrum at 1~cm.  

{\bf G5.89-0.39} contains an ultracompact source (Source A) with broad
RRL emission.  Zijlstra et al. (1990) detected a normal ($\Delta$V= 22 \kms)
and a broad ($\Delta$V= 64 \kms) H76$\alpha$ component toward this source.
Their continuum measurements imply a spectral index of $\nu\uc{2.2}$
from 20~cm to 6~cm and $\nu\uc{0.8}$ from 6~cm to 2~cm.
Afflerbach et al. (1996) measured line widths of 59 \kms\ and 61 \kms\ at
H42$\alpha$ and H66$\alpha$, respectively.  Their data imply a spectral
index of $\nu\uc{0.55}$ from 1.3~cm to 3~mm.
Additional continuum data in the literature give somewhat inconsistent 
results (Acord, Churchwell, \& Wood 1997; Acord, Walmsley, \& Churchwell 1998;
 Wood and Churchwell 1989), but all
of the results imply a rising spectrum throughout the centimeter band
and continuuing to at least 7~mm. Acord et al. (1998) have used 
3.6 ~cm continuum observations over a 5 year span to measure what
they argue is a proper motion of a shell due to dynamical expansion.

{\bf MWC~349} is an isolated HII region in  Cygnus.  
This object is the archetypal
 ionized stellar wind source. Its spectrum rises as
$\nu^{0.6}$ from cm wavelengths  up to 60 $\mu$m (Harvey, Thronson, \& Gatley
1979).
This behavior is characteristic of  
an isothermal wind expanding at constant velocity (see e.g. Wright \& Barlow 
1975). 
The typical 
line widths measured for the centimeter wave recombination lines
(Altenhoff et al. 1981) and optical [N~II] lines (Hartmann, Jaffe, \&
Huchra 1980)  is $\sim$ 60 \kms. VLA 
observations of the H66$\alpha$ line show that  the lines have nearly 
flat topped profiles with a width of around 100 \kms, 
substantially  larger than 
those measured from the Gaussian profiles of the 3 mm  recombination lines 
(Mart{\i}n-Pintado et al 1989). Modeling of the continuum and the line 
intensities of the H66$\alpha$ lines  indicates that the discrepancies are 
likely to be due to non-LTE effects in the centimeter recombination lines (see 
Mart{\i}n-Pintado et al 1993). 
MWC~349 shows a bipolar morphology (White \& Becker 
1985), but unlike K3-50A (see below),  
it does not show a bipolar distribution in the 
recombination lines (Mart{\i}n-Pintado et al 1993); 
indicating that MWC~349 must 
be oriented perpendicular to the line of sight.
As an object for study, MWC~349 has the advantage over the other BRLOs
detected so far that it
is isolated and the ionizing star can be 
seen in the visible and 
the near-IR.
In some sense, MWC~349 may play the role for massive stars that field
TTauri stars play in studies of low mass stars.

At 2--6 cm, the elongated continuum source K3-50A dominates the 
radio continuum emission from compact radio sources in the K3-50 complex.
Combining the  continuum data of Roelfsema, Goss, \& Geballe (1988) and De Pree
et al. (1994), we estimate a 5-15 GHz spectral index for this source
of $\nu\uc{0.5}$.  Roelfsma et al. (1988) detected a broad 
($\Delta$V= 66 \kms) component to the H110$\alpha$ line but attributed
its breadth to pressure broadening.  The more recent H76$\alpha$ results
(De Pree et al. 1994) show not only that the large line widths are
a dynamical effect but also that there is clear evidence in this source
for an ionized bipolar outflow.  Toward the center of the source,
the 2~cm recombination line has a width of 76 \kms.  Three arc seconds  north
and south of this position, the width is only 35 \kms\ but the velocity
changes from -11 \kms\ north of the peak to -42 \kms\ to the south.

\section{Discussion}

\subsection{Observational Constraints on Possible Models}

In this section, we summarize the observations and key parameters
derived from the observations for which any viable model for
objects that produce the broad recombination lines  
(BRLOs) must account.

\subsubsection{ Lifetimes}

BRLOs are common in UC H~II region complexes. The lifetimes
of these sources are much longer than their crossing times and the
broad line emission phase must represent a substantial fraction of the
lifetime of UC H~II sources.

One argument for the long duration of the broad line emission phase
as a part of overall H~II region evolution is the presence of broad
millimeter and high frequency centimeter recombination lines toward
ultracompact sources like
NGC~7538~IRS1 and MWC~349 and also toward more extended and presumably
more evolved sources like 
K3-50  and  S106.    It is possible that there is a connection
between the dominant broad line emission at early evolutionary stages
and the low-level emission at high velocities seen in very
evolved HII regions (Meaburn \& Walsh 1981, 1986).   
The available sample of H~II regions observed
with high enough sensitivity to detect the broad recombination 
lines is small (Table \ref{tab3}).
As a consequence, our estimate of the fraction of compact  H~II 
regions with   broad  lines 
is somewhat uncertain and may suffer from selection biases. 
We follow two approaches in estimating the frequency 
of occurrence of  these objects: We first 
determine  the fraction of sources in UC H~II region  complex that  show the 
broad lines, 
within each well-studied complex containing at least one BRLO.
In NGC 7538, one out of 
three sources is a BRLO. In G34.3+0.2 one to three out of three is likely to be
a BRLO and in W49, we find that two out of eight sources are probably
responsible for the broad recombination line emission.
From these data,  we estimate that roughly around 
30\% of the compact sources in these complexes are BRLOs. 

As a second approach, we use a limited sample of compact H~II regions
(Afflerbach et al. 1996). Since we are only interested in the statistical
properties of the sample and since the signal to noise in the recombination
line data from this survey is lower than for any of the observations of
individual sources, we set a somewhat lower threshold for BRLOs and
include objects with RRL widths $\geq$45 \kms.
Using this criterion, 6 of the 23 sources in the sample have lines broad
enough to demonstrate the presence of one or more BRLOs
which also implies a frequency of incidence of roughly 30\% for BRLOs
in their sample. 

Table \ref{tab3} gives a
summary of the  main characteristics of  all the well-studied BRLOs
(i.e. not including the statistical sample from the Afflerbach
et al. survey).
Most of the sources are very compact, with typical crossing times
of less than 
2$\times 10^3$ years for expansion velocities of around 50 \kms. 
Following the same line 
of argumentation used by  Wood \& Churchwell (1989) 
for all compact HII regions, 
we find that, for such  short dynamical  times, one would expect to detect only 
$\sim$15  objects of this kind in the whole galaxy, close to the number of 
objects so far detected in just a  few regions. 
The high detection rate for BRLOs means that  
these objects cannot be expanding at the velocities measured by the 
recombination lines. 

\subsubsection{ Continuum Properties}

Most of the resolved BRLOs  have a bipolar morphology and 
almost all BRLOs have rising continuum 
spectra up to at least 1.3 cm (Table \ref{tab3}). Even for the most compact objects, the 
rising spectrum continues well into the millimeter wavelengths. 
Unfortunately, the 
complexity of most of the regions and the presence of large column densities 
of dust  makes it very difficult to measure the continuum spectral indexes at 
short millimeter and sub-millimeter wavelengths. For  the  best case, MWC~349, 
the $\nu^{0.6}$ continuum spectrum is observed up to 60 $\mu$m (Harvey
et al. 1979). If other BRLOs share this characteristic,
large emission measures lie hidden in the optically thick regions
seen at centimeter
wavelengths in most of the BRLOs in Table 3. As determined from the 
high-resolution VLA maps, all the  BRLOs have emission 
measures of at least 10$^8$ 
pc cm$^{-6}$ and the number of the Lyman continuum photons
needed to explain the radio 
data varies from 10$^{47.6}$ s$^{-1}$ for S106 to more than   10$^{48.6}$ 
s$^{-1}$ for the other sources. For the sources with  continuum spectra 
that rise into the millimeter
wavelengths, these values are clearly lower limits and the emission 
measure and the number of Lyman continuum photons increase with frequency at 
least with  the same power law as the continuum spectrum. 
This indicates that the ionization  of all the BRLOs, except for S106, 
will require stars with spectral types earlier 
than O6. The large emission measure indicates that the 
BRLOs must be  ionized by a star or stars at the  geometrical centers of 
the HII regions, within the optically thick cores observed at 
centimeter wavelengths.  The central location of the ionizing stars is
already apparent in MWC~349 and S106 
where the geometry is clear because the ionizing stars are observed in the visible and the IR. 

The rising continuum spectral index of an opaque ionized source
is a product of  a combination of the radial  
density and temperature distributions of the ionized material. 
This index is nearly independent of the axial and azimuthal
geometry (Schmid-Burgk 1982). The BRLOs mapped with the VLA 
at high angular resolution show that the continuum and recombination line 
intensities are consistent with minor variations of the electron temperatures 
within the sources (see e.g De Pree et al. 1994 for K3-50, 
Gaume et al. 1995 for 
NGC 7538 IRS1; Mart{\i}n-Pintado et al. 1993), suggesting that the HII 
regions are 
close to isothermal. Furthermore, for the typical spectral indexes measured for 
the BRLOs,  between 0.5 and 1.5, the effects of moderate power law temperature 
gradients have  minor effects ($\sim$20\%)  
on the  continuum spectral  indexes; 
being independent of temperature gradients for spectral indexes of  $\sim$0.6 
(Cassinelli \& Hartmann 1977). Therefore, we will consider that the spectral 
indexes given in Table 3 are basically related  to the density structure within 
the BRLOs. The density structure varies from 
n$\propto$r$^{-2}$ to n$\propto$r$^{-3}$ for  
spectral indexes between $\simeq$0.6 and 1.3 respectively
(Olnon 1975; Panagia \& Felli 
1975; Wright \& Barlow 1975). For such steep density gradients, the HII region 
is far from pressure equilibrium and the ionized material must be 
flowing away from the dense region near the star. 
From mass conservation, the density  
structures derived from the 
continuum spectral indexes, imply that only in a small fraction of BRLOs 
(those with spectral indexes of $\sim$ 0.6) is the ionized gas outflowing at 
constant velocity, and in most of the BRLOs the ionized gas is accelerating.
In the most extreme case 
(spectral indexes of 1.3),  the velocity must be 
proportional to the radius. 

\subsubsection{ Line Widths}

The measured linewidths of the BRLOs (Table \ref{tab2}) fall within a
remarkably small range.  Our best estimate for the completeness range
for our data is $\Delta$V(FWHM)=45--100 \kms.  Confusion with narrow
line emission from the same complex imposes the lower bound.  We
estimate the upper bound from our observed spectra.  The upper bound
results from a combination of the diminuition of peak intensity with
increased width and confusion with the helium lines and with 
low-level molecular line emission.
From our own observations, however, the widths of the broad recombination
line components cluster strongly around 75 \kms.  The results in
Table \ref{tab3} show that even sources observed at centimeter wavelengths
with interferometers have broad recombination lines with widths
very close to 75 \kms.  If further observations bear out this 
consistency in the velocity spread of the emission, the explanation
of this result must form an important part of any theory of BRLOs.

\subsection{Possible Models}

In this section, we summarize the various possible models for UC H~II
regions in the context of the our new data and  other
recently published data on broad recombination line emission.
We note first  that the BRLOs are the products of gas 
dynamics, rather than radiative transfer.
RRL observers are conditioned to think about pressure broadening
whenever they see broad line profiles or wide wings on the lines.
We can rule out pressure broadening, however, as the source of the wide
lines in BRLOs.
The very strong dependence of the pressure broadened line width on
principle quantum number, n ($\Delta$V$\propto$n$\uc7$) means that,
if pressure broadening produces the wide lines,
their widths should change dramatically (by factors of 5-400)
over the range of n values observed for our BRLOs.  
In all cases for which there are observations of
Hn$\alpha$ transitions over a range of n, the {\it observed} changes in the 
width and/or strength of the broad recombination line component
are much smaller than those predicted by pressure broadening.

\subsubsection{ Champagne Flows}

Champagne models (Tenorio-Tagle 1979, Bodenheimer, Tenorio-Tagle, \&
Yorke 1979)
consider the evolution of young H~II regions as they break out of the
dense neutral cores in which they form into a lower density diffuse medium.
During the breakout phase, the ionized gas accelerates into the low
density medium at speeds up to 30--50 \kms.  Subsequent analysis shows 
that this kind of acceleration also occurs in H~II regions that form in clouds
with power law density distributions, if the density gradients are
sufficiently steep (Franco, Tenorio-Tagle, \& Bodenheimer 1990).  
A champagne phase should be a
common occurence among UC H~II regions, if they really form as a result
of ionization of the surrounding molecular core, because of their
extremely large internal pressures.


\subsubsection{ Ionization of Neutral Winds}

Hypersonic bipolar outflows of molecular gas are a common feature of 
protostellar sources, both in low mass star forming regions and in 
regions containing low and high mass pre main sequence objects
(Bachiller 1996; Plume et al.  1992).  
These outflows have velocities ranging from a
few up to about 100 \kms.  External ionization of molecular flows
near O stars embedded in a young cluster could reproduce at least
some of the features of BRLOs.  Alternatively, if the turn-on time 
for the ionizing star is short enough and occurs soon enough after
the outflow phase, the O star could ionize its own molecular flow from
the inside.

K3-50A is the only one of the broad line sources that is large enough
(0.2$\times$1.2 pc) to have a reasonably long crossing time 
($\sim$2$\times 10\uc4$ years).  In principle, then, this source
would be a good candidate for the ionization of a fossil flow.
In such a picture, we would expect the ionizing photons
to move through the cavity evacuated by a fast molecular wind
and strike the swept-up material in the cavity walls.  
The dense, UV-illuminated walls would appear as an elongated, limb-brightened
shell.  Existing continuum maps do not show such a shell (De Pree et al. 1994).

\subsubsection{ Wind-Driven Bubbles}

The fast (1000-2000 \kms) stellar winds from young O stars
will interact strongly with the surrounding interstellar gas.  After a
short time ($\sim$10$^3$ years, Shull 1980, 1982), the wind will create
a bubble with a dense shell of swept-up material.  If this bubble forms
in a flattened structure, it may result in formation of a bipolar
outflow of ionized material (Koo \& McKee 1992).

\subsubsection{ Mass-Loaded Winds}

Redman, Williams, \& Dyson (1995, 1996; Dyson, Williams \& Redman 1995)
and Lizano et al. (1996) have developed a scheme to produce
long-lasting UC H~II regions in which material photoevaporated from
dense circumstellar clumps loads the O star wind and produces a dense
outflow of ionized material bounded by a recombination front.

\subsubsection{ Disk Winds}

Hollenbach et al. (1994) and Yorke and collaborators (Yorke \& Welz 1993, 1996;
Richling \& Yorke 1997) have suggested that evaporative flows from massive
circumstellar disks may offer a solution to the lifetime problem for
UC H~II regions.  In their models, stellar UV photons ionize material
in the disk and produce steady mass loss for periods of up to 10$^5$ years.
These flows are inherently bipolar.  It is possible, however, to produce
asymmetric regions by invoking external photoevaporation of the disks of
nearby companion stars (Kessel, Yorke, \& Richling 1998).
 Most of the papers discuss
the terminal velocity of this
evaporated material but do not explicitly model it. Kessel et al. (1998)
present line profiles for optical transitions showing typical linewidths
on the order of 40 km s$^{-1}$.

\subsubsection{Remarks on the Models}

Since the widespread existence of BRLOs was unknown at the time the various
theoretical papers were published, the models do not explicitly address
the issues raised by the nature of these sources.  Nevertheless, the
papers do allow us to categorize the potential problems of the various models.
None of the models has an obvious explanation for the relatively narrow
range in observed line widths. 
All but the last two models suffer from difficulties with the lifetime.
They produce high velocity emission only for the order of a single
crossing time and therefore cannot account for the number and frequency of
observed BRLOs.  We therefore consider only the 
mass loaded wind and photoevaporated disk models in detail.

 The mass-loaded wind models can only account for the
bipolar morphology of most BRLOs if the underlying protostellar wind is
strongly bipolar (the photoevaporated disk models, by contrast, only work
if {\it all} of the BRLOs turn out to be bipolar).  Mass loaded wind
models also fail to explain the r$^{-2}$ or steeper density gradients
inferred from the continuum spectral indices.  Mass loading should result
in density profiles less steep than r$^{-2}$. 
The existing versions of the mass-loaded wind and 
photoevaporated disk models also do not predict the 
very broad recombination line profiles we observe.
 
On balance, the photoevaporated disk model has fewer problems than the 
mass-loaded wind model.
 There are observations that provide
some supporting observational
evidence for the presence of disks, if not for their
involvement in the generation of the broad lines.
For the best-studied BRLO, MWC~349, the  hot dust 
surrounding the star shows  a disk-like distribution with a size of  
$\leq$50x100 AU (Leinert  1986). A similar size for the neutral disk has also 
been inferred from the high angular resolution radio continuum map at 1.3 cm 
(Mart{\i}n-Pintado  et al. 1993). Direct evidence of the interface region 
between the neutral disk and the ionized gas  has been measured from the 
H30$\alpha$  recombination line maser spikes  which also indicates a face-on 
disk of  $\sim$ 80 AU (Planesas, Mart{\i}n-Pintado \& Serabyn 1992). The most 
detailed knowledge of the kinematics of the circumstellar gas comes from the IR 
recombination lines (Hamann \& Simon 1986) and the observations of the 
H30$\alpha$ maser line. Modeling of the IR line profiles indicates the presence 
of a neutral rotating disk and a bipolar ionized wind. The location and the 
radial velocities of the  H30$\alpha$ maser spikes are consistent with a 
keplerian disk rotating around a 30 M$_{\sun}$ 
object (Planesas et al. 1992; 
Thum, Mart{\i}n-Pintado \& Bachiller 1992). The  presence of a 
disk in S106 is also inferred from the morphology of the ionized gas (see Felli 
et al. 1985). Unfortunately, for the other BRLOs,
 the properties of the disks are unknown. We note, however, that these 
sources are 
excellent candidates for the detection of recombination line masers at 
millimeter  wavelengths, similar to those observed in 
MWC~349 (Mart{\i}n-Pintado et al. 1989). Detection of maser emission might
allow us to observe the 
interface region between the neutral disk and the ionized material.

\subsection{Conclusions}

1) BRLOs are common in regions containing embedded OB stars.  These
objects have recombination line widths of 60--80 \kms\ and continuum
spectra that rise as $\nu^{\sim0.3}$ to $\nu^{\sim1.0}$ into the 
millimeter band.  Most spatially resolved BRLOs have bipolar morphology.

2) The short crossing times and high frequency of occurence of BRLOs
implies both that these objects last for much longer than their expansion
times and that many UCH~II regions must begin their lives as BRLOs.

3) The high continuum opacity of BRLOs at short centimeter wavelengths
means that estimates of dust extinction based on near-IR lines and radio
continuum measurements probably overestimate the extinction.

4) None of the published models of UCH~II regions explicitly deals
with all of the observed properties of BRLOs.  If a photoevaporated
disk is the source of the
ionized bipolar emission region, the results imply that disks around
massive young stars are both common and long-lived.

This work was supported in part by National Science Foundation grant 95-30695
to the University of Texas at Austin.

\vfil
\eject
\centerline{\bf Figure Captions}
\figcaption{ \label{n7538comp}
Comparison of fits of one and two Gaussians to the H41$\alpha$ line
toward NGC~7538~IRS1. ({\it left}) Single Gaussian fit to the region
within $\pm$15 \kms\ of the peak intensity and residuals to that fit.
({\it right}) Two Gaussian fit to the same profile (line parameters
in Table \ref{tab2}).  Most of the power in the residuals to this fit is due
to either He41$\alpha$ or to the molecular feature seen at higher velocities
(CH$_3$CN).
}

\figcaption{ \label{w49n}
Comparison of fits of one and two Gaussians to the H41$\alpha$ line toward
W49N. ({\it upper left}) Single Gaussian fit to the region within
$\pm$15 \kms\ of the peak intensity. ({\it lower left}) Residuals to the
single Gaussian fit. ({\it upper right}) Two Gaussian fit  to the same
profile (see Table \ref{tab2}). ({\it lower right}) Residuals to the
two Gaussian fit.  The solid line shows a two Gaussian fit to the He41$\alpha$
line made using the widths of the hydrogen features and a constant velocity
offset of --122.2 \kms.
}

\figcaption{ \label{w49compadd}
Comparison of H39$\alpha$ (thin histogram) and H41$\alpha$ (dashed histogram)
profiles in W49N. ({\it upper panel}) The dotted line shows the third 
feature in a three Gaussian fit to H39$\alpha$.  For this fit, the width
and position of the other two components were fixed to the values derived
from a fit of two Gaussians to the H41$\alpha$ profile. ({\it lower panel})
Overlay of the H39$\alpha$ profile and an H41$\alpha$ profile to which
the third component has been artificially added.
}

\figcaption{ \label{w44}
Comparison of fits of one and two Gaussians to the H41$\alpha$ line toward
G34.3+0.2. ({\it upper left}) Single Gaussian fit to the region within
$\pm$15 \kms\ of the peak intensity. ({\it lower left}) Residuals to the
single Gaussian fit. ({\it upper right}) Two Gaussian fit. 
({\it lower right}) Residuals to the two Gaussian fit and a two Gaussian fit
to the He41$\alpha$ line (see text).
}

\figcaption{ \label{w44comp}
Overlay of the H39$\alpha$ (solid line) and H41$\alpha$ (dashed line)
spectra toward G34.3+0.2.  The dashed line shows the sum of two Gaussians
representing a compromise between fits to the two hydrogen transitions
(V$_N$= 53.0 km s$^{-1}$, $\Delta$V$_N$=32.3 km s$^{-1}$; V$_B$=41.6 km s$^-1$,
$\Delta$V$_B$=75.0 km s$^{-1}$).
}

\figcaption{ \label{oria}
({\it left}) H39$\alpha$ toward Orion H$_2$ Peak 1.  The smooth curve shows
a single Gaussian fit to the observed spectrum (see Table \ref{tab2}).  The
residuals to this fit are below the profile. ({\it right}) H41$\alpha$
toward Orion IRc2.
}

\figcaption{ \label{g106}
Recombination line spectra toward G10.6-0.4. ({\it top}) H41$\alpha$.
The smooth curve shows a single Gaussian fit to the hydrogen line and a
fit for the intensity of the He41$\alpha$ line with the width of this
line fixed to the hydrogen line width and the velocity fixed to -122.2 \kms\
with respect to the hydrogen line. ({\it bottom}) Fit to the H35$\alpha$
line and He35$\alpha$ line.
}

\figcaption{ \label{dr21}
Single Gaussian fit to the H39$\alpha$ line toward DR21.  The He39$\alpha$
line was fit holding the width equal to the H39$\alpha$ width and the velocity
at --122.2 \kms\ with respect to the hydrogen line.
}
\vfil
\eject


\begin{deluxetable}{ccccc}
\footnotesize
\tablecaption{Observing Log \label{tab1}}
\tablewidth{0pt}
\tablehead{
\colhead{Source}                      &
\colhead{R.A.(1950)}                  &
\colhead{Decl.(1950)}                 &
\colhead{Date} 		              &
\colhead{Lines}    
}
\startdata
Orion H$_2$ Peak 1 & 05 32 46.3 & -05 23 54 & 2/92 & H30$\alpha$, H36$\alpha$, H39$\alpha$ \nl
Orion IRc2 & 05 32 47.0 & --05 24 24 & 2/92 & H36$\alpha$, H41$\alpha$  \nl
G10.6-0.4 & 18 07 30.6 & --19 56 28 & 3/93 & H35$\alpha$, H41$\alpha$ \nl
G34.3+0.2 & 18 50 46.1 & 1 11 10 & 2/92 & H36$\alpha$, H39$\alpha$, H41$\alpha$ \nl
    &            &         & 3/93 & H30$\alpha$, H31$\alpha$, H35$\alpha$, H39$\alpha$, H41$\alpha$ \nl
W49N & 19 07 49.8 & 9 01 17 & 2/92 & H36$\alpha$, H39$\alpha$, H41$\alpha$ \nl
     &            &         & 3/93 & H30$\alpha$, H31$\alpha$, H35$\alpha$, H41$\alpha$ \nl
S106 & 20 25 32.8 & 37 12 44 & 2/92 & H36$\alpha$, H39$\alpha$ \nl
     &            &         & 3/93 & H35$\alpha$, H41$\alpha$ \nl
DR21 & 20 37 14.1 & 42 08 55 & 2/92 & H30$\alpha$, H36$\alpha$, H39$\alpha$ \nl
     & 20 37 14.2 & 42 09 06 & 3/93 & H35$\alpha$, H41$\alpha$ \nl
NGC 7538 & 23 11 36.5 & 61 11 49 & 2/92 & H30$\alpha$, H36$\alpha$, H39$\alpha$ \nl
     &            &         & 3/93 & H31$\alpha$, H35$\alpha$, H39$\alpha$, H41$\alpha$ \nl
\enddata
\end{deluxetable}

\vfil
\eject
\begin{deluxetable}{ccccccccc}
\footnotesize
\tablecaption{Continuum and recombination Line Parameters \label{tab2}}
\tablewidth{0pt}
\tablehead{
\colhead{Source}                      &
\colhead{Line}                        &
\colhead{}                            &
\colhead{T$_L$}                       &
\colhead{V}                           &
\colhead{$\Delta$V} 		              &
\colhead{$\int$TdV}        &
\colhead{S$_\nu$ \tablenotemark{a)} }     &
\colhead{T$_e$$^*$ \tablenotemark{b)}}  \\ 
 \colhead{}                            &  
\colhead{}                            &  
\colhead{}                            & 
\colhead{ K }       &
\colhead{ \kms}   &  
\colhead{ \kms}   & 
\colhead{K  \kms }  &
\colhead{Jy}          &
\colhead{ K}  
}
\startdata
Orion H$_2$ Peak 1 & H39$\alpha$ &  & 0.27 & --4.34 & 25.6 & 7.3(0.25) &  &   \nl
Orion IRc2 & H41$\alpha$ & & 0.54 & --3.57 & 22.7 & 13.0(0.3) &  &    \nl
G10.6-0.4 & H41$\alpha$ & & 0.67 & 0.2(0.1) & 28.0(0.1) & 20.5(0.15)& 4.0&9300 
\nl
G34.3+0.2 & H41$\alpha$ & N & 0.45 & 53.0 & 32.3 & 14.9(0.4) & 3.0 & 9500 \nl
          &             & B & 0.25 & 41.6 & 75.0 & 21.1(0.6) & 4.2 & 9500\nl
W49N & H41$\alpha$ & N & 0.77  & 7.7  & 30.6 & 25.1 & 6.0 & 10500 \nl
     &             & B & 0.18  & 12.7 & 70.8 & 13.3 & 3.0 & 10500\nl
S106 & H41$\alpha$ & N & 0.136 & --3.1(0.6) & 29.7(2.3) &  4.3(0.7) & 1.0 & 
10000 \nl
     &             & B & 0.066 & --1.6(2.2) & 75(8)     &  5.2(0.8) & 1.1 & 
10000 \nl
DR21 & H41$\alpha$ &    & 1.24  & --0.84(0.06) & 32.6(0.1) & 43.2(0.2) & 10.6 & 
11000\nl
NGC 7538 & H41$\alpha$ & N & 0.094 & --65.6(0.7) & 24.5(1.0) & 2.7(0.3) &0.8 
&12800 \nl
         &             & B & 0.077 & --47.8(1.9) & 69.3(3.0) & 5.8(0.4) &1.7 
&12800 \nl
        & H39$\alpha$ & N & 0.095 & --65.6(0.6) & 26.8(1.9) & 2.7(0.3) & 
& \nl
         &             & B & 0.066 & --46.8(2.1) & 77.1(3.9) & 5.4(0.3) & 
&
\enddata
\tablenotetext{a}{The continuum flux densities for 
the normal (N) and broad (B) spectral components have been 
derived from the electron temperatures and the integrated line intensities}
\tablenotetext{b}{LTE electron temperature derived from the ratio between the 
total integrated line intensity (B+N)  and the measured continuum emission}

\end{deluxetable}

\vfil
\eject



\begin{deluxetable}{ccccccc}
\footnotesize
\tablecaption{ Summary of the properties of the broad recombination line sources 
\label{tab3}}
\tablewidth{0pt}
\tablehead{
\colhead{Source}                      &
\colhead{Morphology}                        &
\colhead{Spectral}                            &
\colhead{Size}                       &
\colhead{$\Delta$V}                           &
\colhead{Cross. time }        &
\colhead{References to}  \\ 
\colhead{ }                            &  
\colhead{ }                            &  
\colhead{ index }                            &  
\colhead{ pc }                            & 
\colhead{ kms$^{-1}$}       &
\colhead{ 10$^2$ yr}   &  
\colhead{col. 2 to 5 } 
}
\startdata
Sgr B2F$^*$ & Compact & $\sim $1.5 & 0.011& 59 & 2 & a \nl
G5.89-0.39 & Shell  & 0.6, 2.2  & 0.044 & 64 & 6 & b, c, d\nl
G34.3+0.2C & Cometary  & 0.9, --0.1 &0.035 &75 &5 & b, e\nl
W49N & Unresolved  &  0.6, 0.9 & 0.07 & 71 & 10 & f, e \nl
K3-50A & Bipolar & 0.5 &0.2x1.2 & 66 &  30-180& g \nl
S106  &   Bipolar & 0.73, --0.1  & 0.07x0.2 &75 &9-12& h, e \nl
MWC349 & Bipolar & 0.65 & 0.001x0.03 &  60  &  0.2-0.6 & i, j, k \nl
NGC7538 IRS1 & Bipolar  & 0.9  & 0.003x0.02 & 64 & 0.5-3& l, e \nl
\enddata
\tablenotetext{}{
$^*$ 4 sources,
a) De Pree et al. (1996), 
b) Wood \& Churchwell (1989), 
c) See text, 
d) Zijlstra et al. (1990),
e)  this paper, 
f) De Pree et al. (1996),
g) De Pree et al. (1994),
h) Bally, Snell \& Predmore (1983),
i) Mart{\i}n-Pintado et al. (1993),
j) Harvey, Thronson \& Gatley (1979),
k) Hartmann, Jaffe \& Huchra, (1980), 
l) Gaume et al. 1995) }

\end{deluxetable}

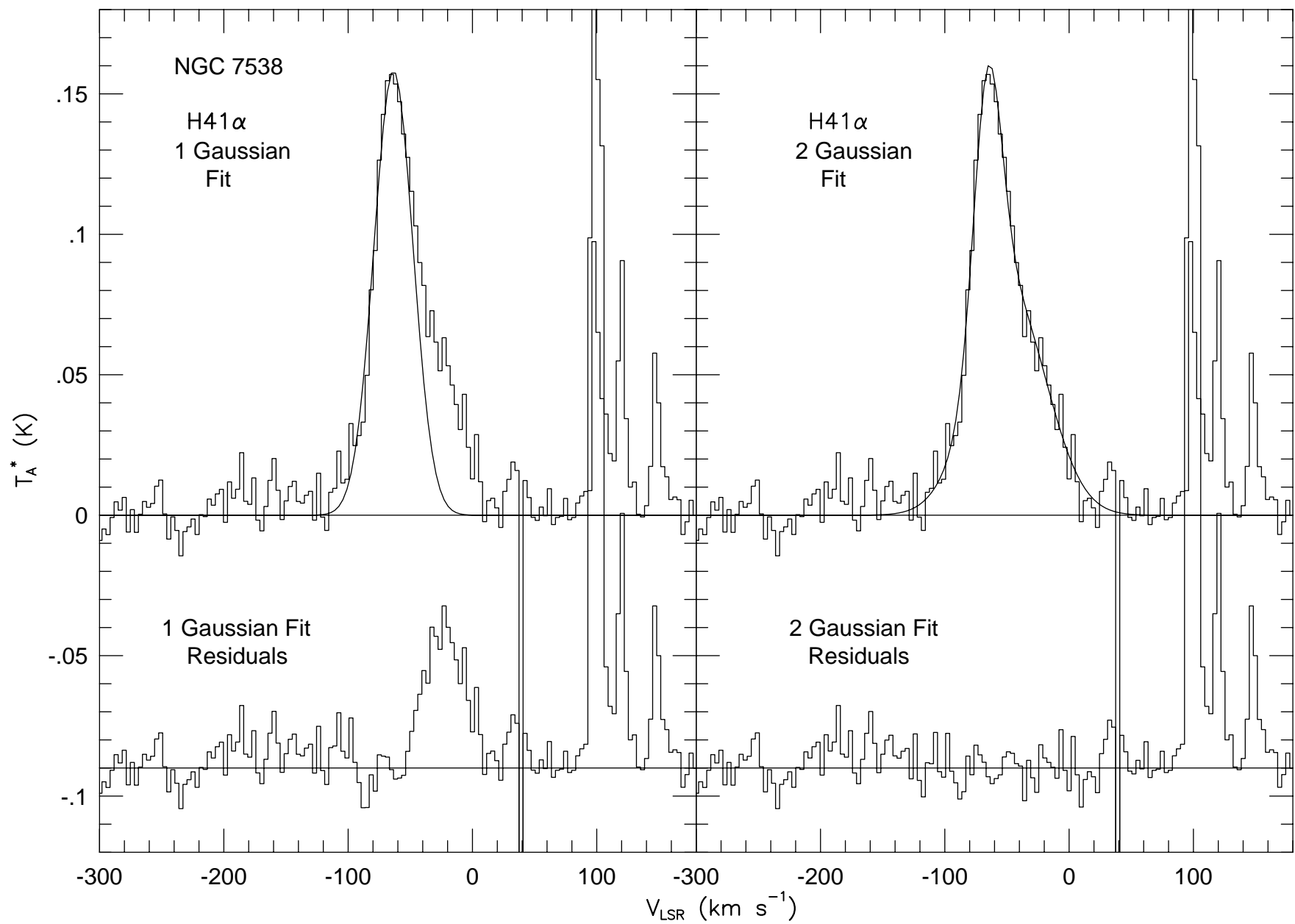
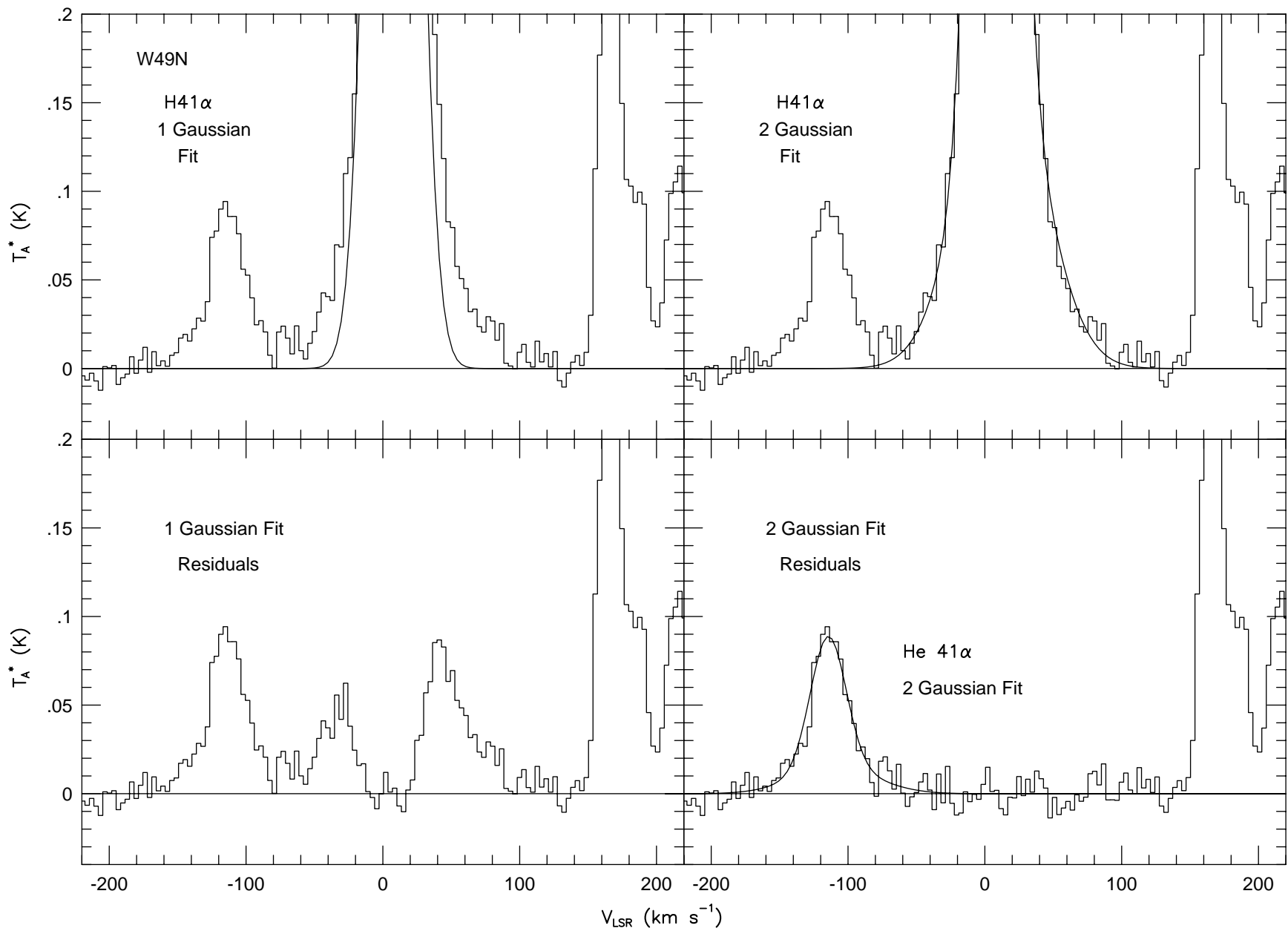
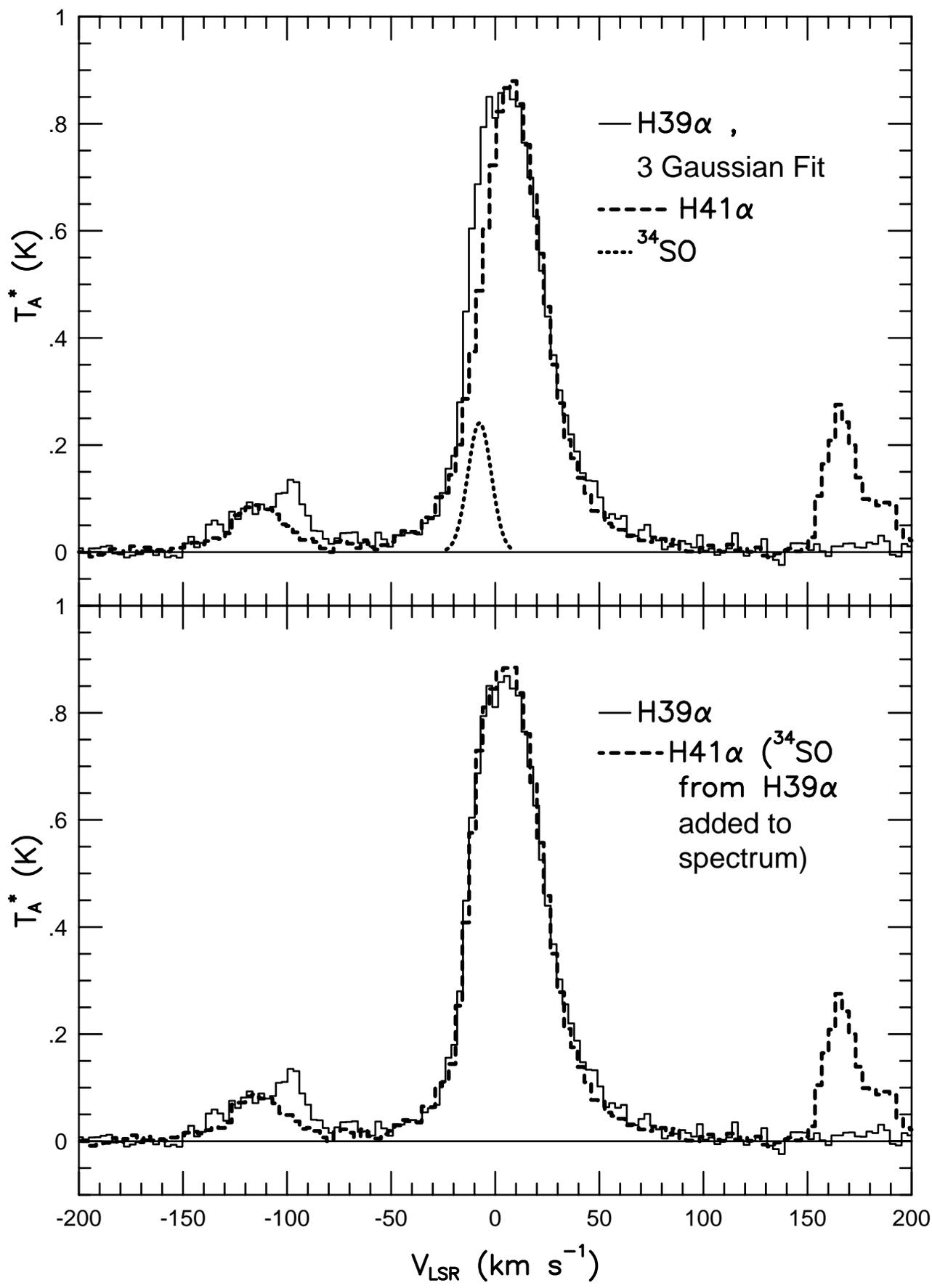
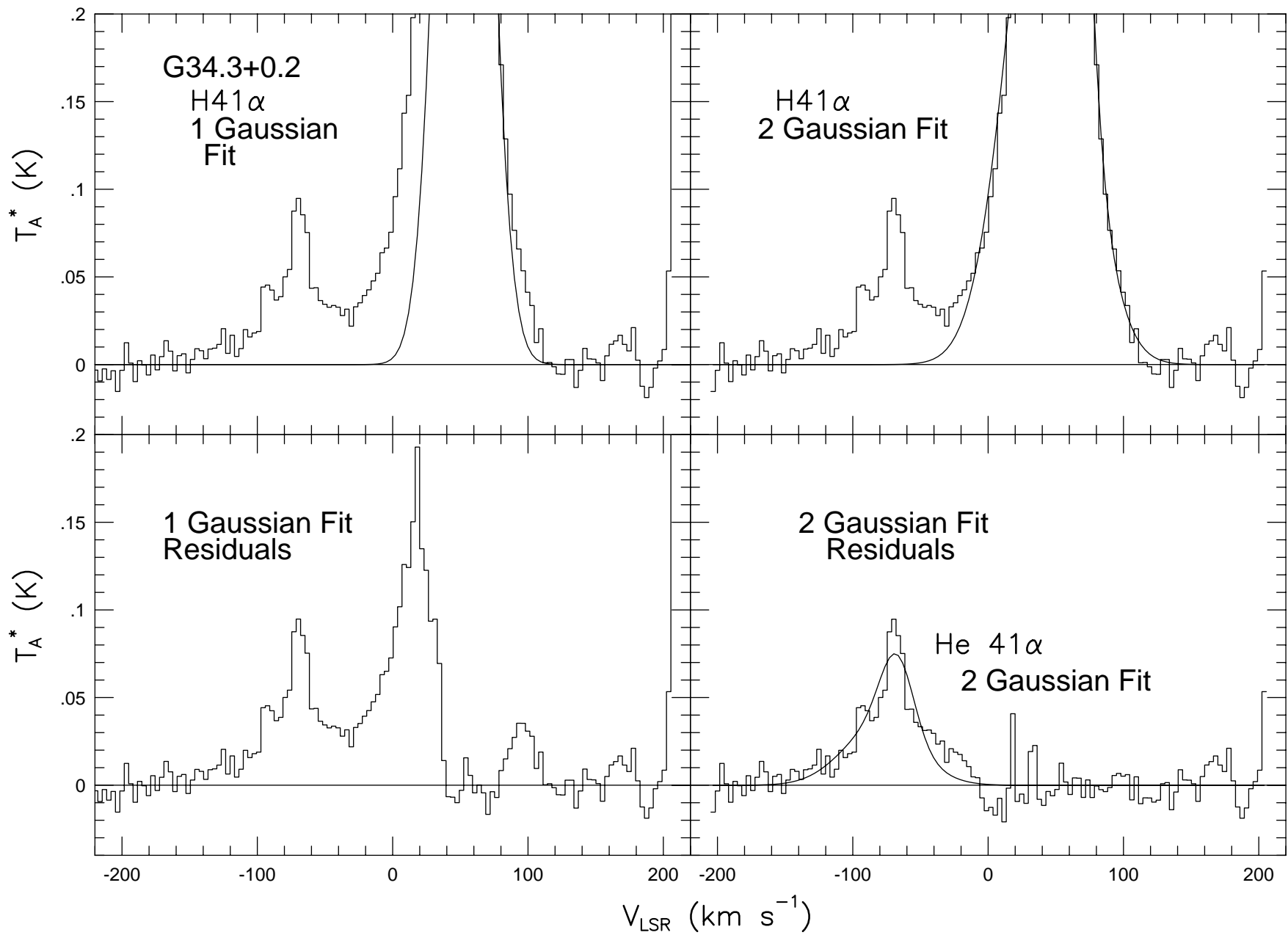
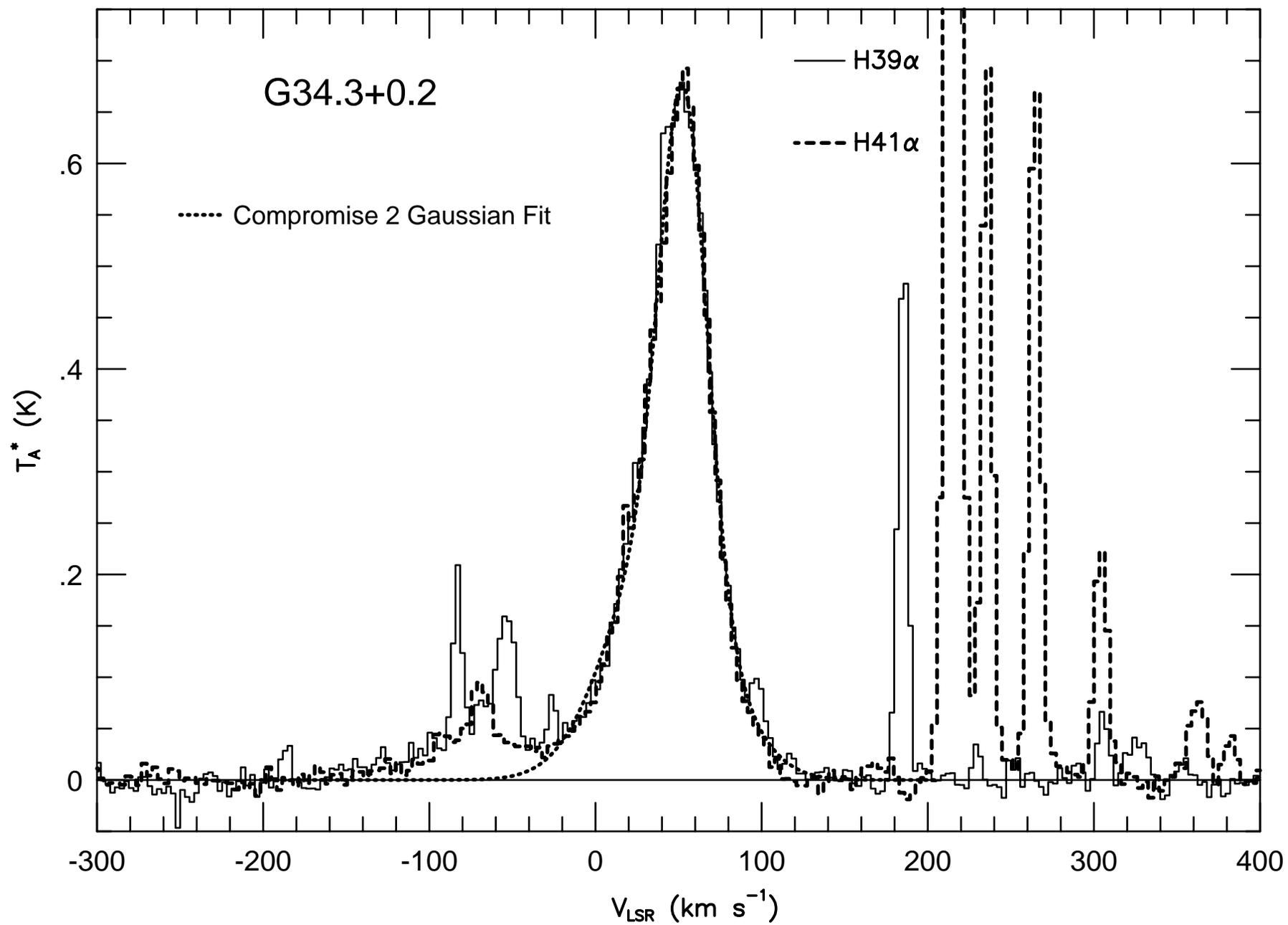
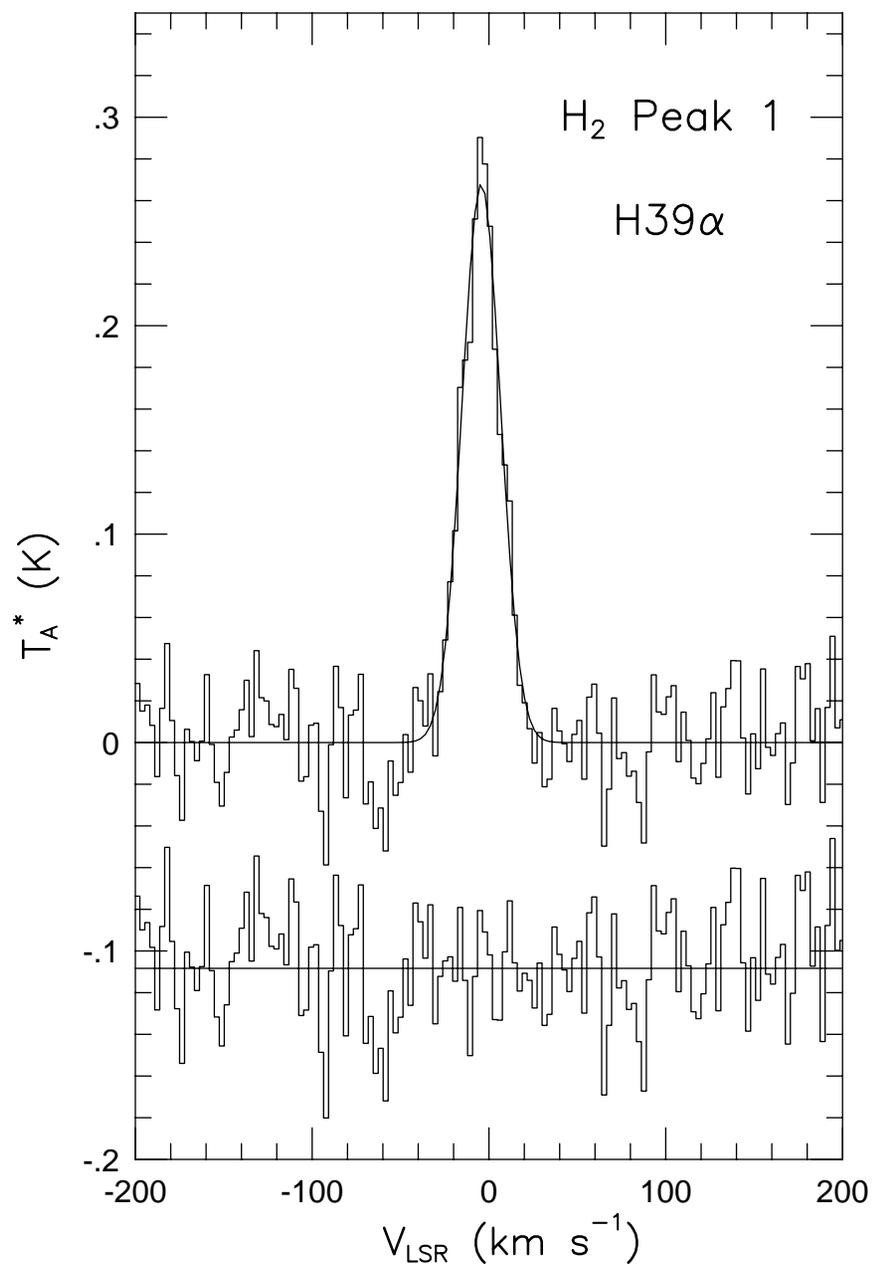
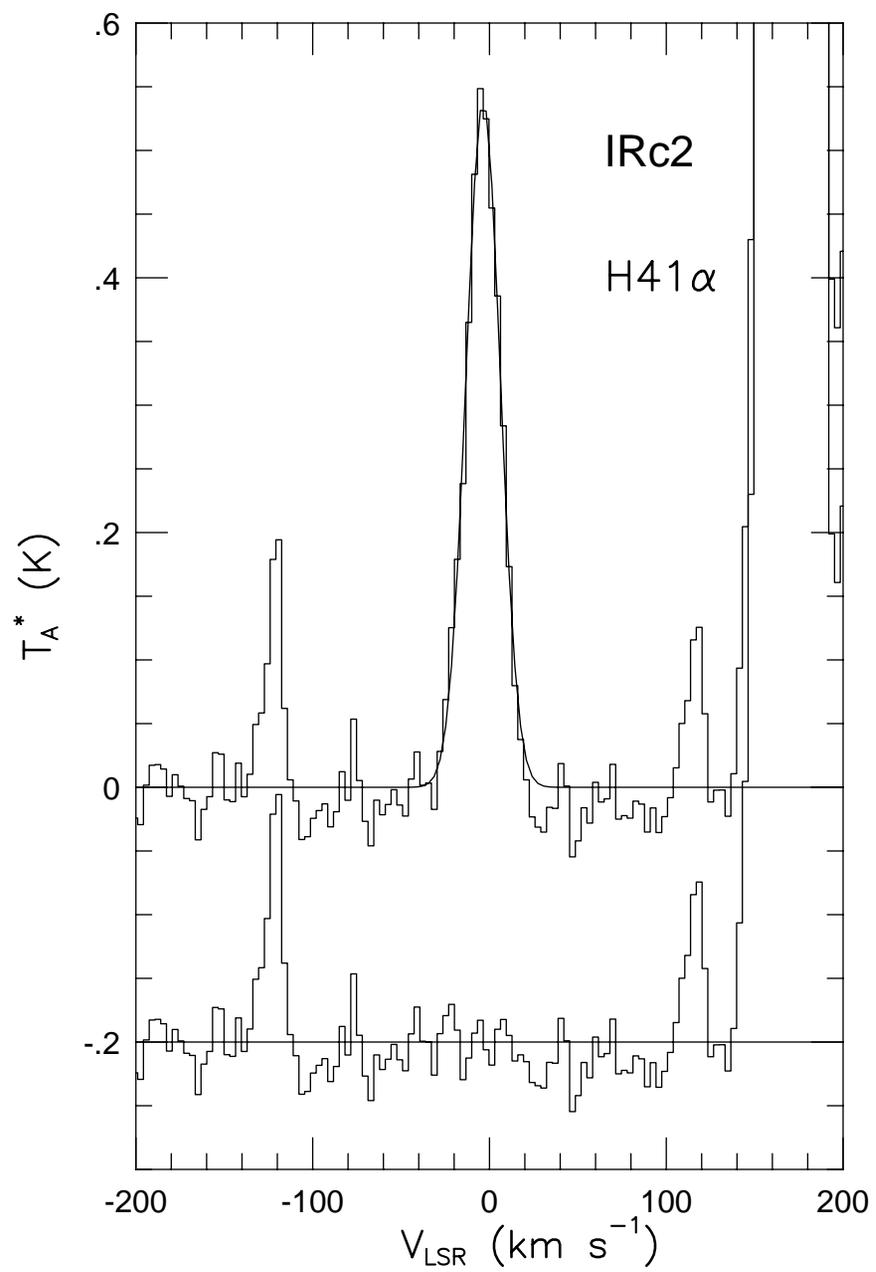
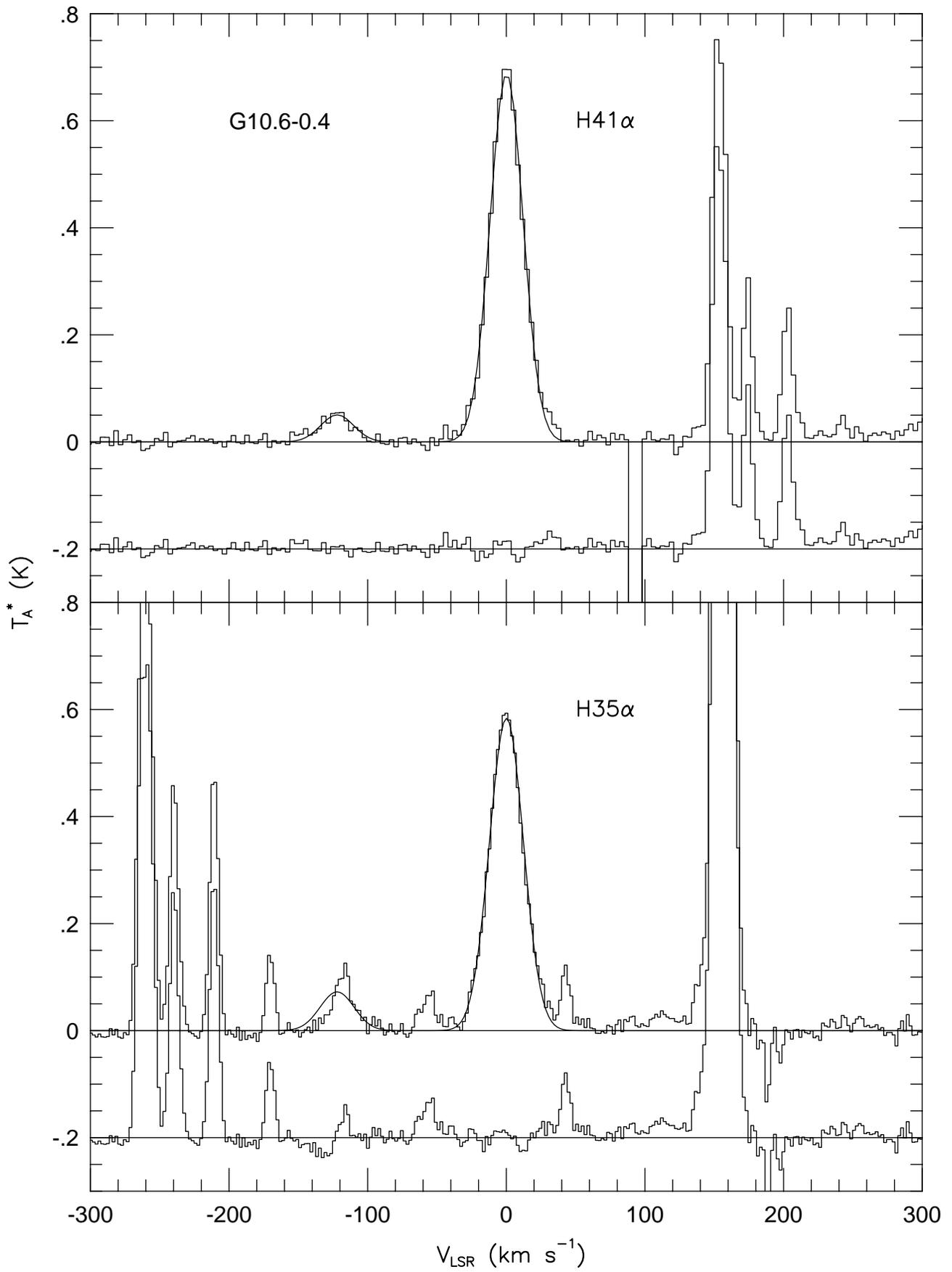
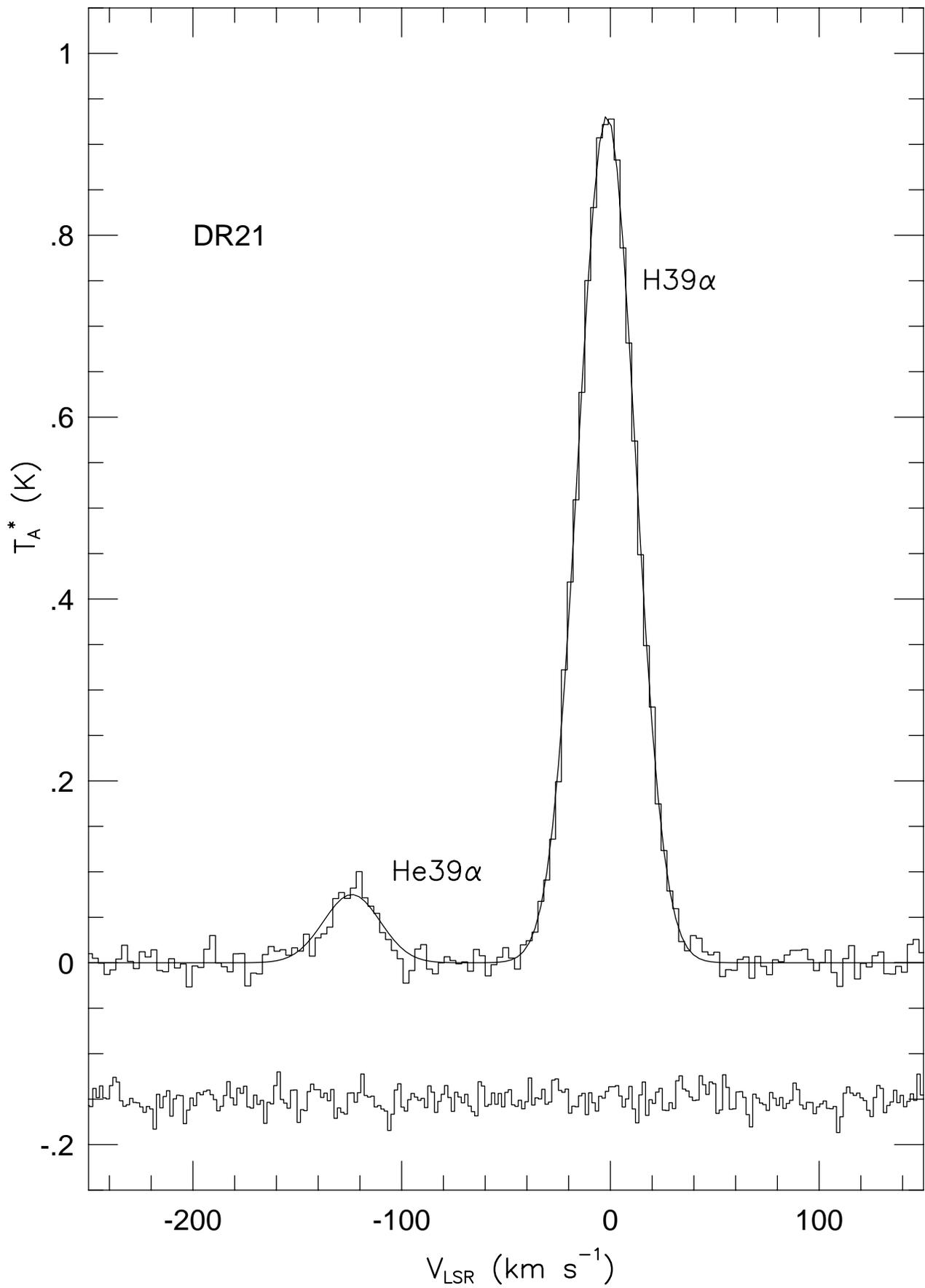
\end{document}